\documentclass{ws-procs9x6}

\usepackage{slashbox}
\def\beq{\begin{equation}}
\def\eeq{\end{equation}}

\begin{document}

\title{Soft Gluon Logarithmic Resummation and Hadron Mass Effects in Single 
Hadron Inclusive Production\footnote{\uppercase{W}ork 
supported in part by \uppercase{D}\uppercase{F}\uppercase{G}
through \uppercase{G}rant \uppercase{N}o.\ \uppercase{K}\uppercase{N}~365/3-1 and by 
\uppercase{B}\uppercase{M}\uppercase{B}\uppercase{F} through 
\uppercase{G}rant \uppercase{N}o.\ 
05~\uppercase{H}\uppercase{T}4\uppercase{G}\uppercase{U}\uppercase{A}/4.}}

\author{S. ALBINO}

\address{2nd Institute for Theoretical Physics, Hamburg University, Germany}

\maketitle

\abstracts{
We define a general scheme for the evolution of fragmentation functions which resums
soft gluon logarithms in a manner consistent with fixed order evolution.
We present an explicit example of our approach in which double logarithms are
resummed using the Double Logarithmic Approximation.
We show that this scheme reproduces the Modified Leading Logarithm Approximation in certain limits,
and find that after using it to fit
quark and gluon fragmentation functions to experimental data, 
a good description of the data
from the largest $x_p$ values to the peak region in $\xi=\ln (1/x_p)$
is obtained. In addition, we develop a treatment of hadron mass effects
which gives additional improvements at large $\xi$.
}

\section{Introduction}

\label{Intro}

The extraction of fragmentation functions (FFs) at large and intermediate
$x$ from experimental data on single hadron inclusive production
has been successfully performed \cite{KKP2000} using
the fixed order (FO) DGLAP \cite{DGLAP} evolution to next-to-leading 
order (NLO) \cite{Curci:1980uw;Furmanski:1980cm}. However,
this formalism fails at small $x$ due to unresummed soft gluon logarithms (SGLs), 
and a different analysis is required in which these SGLs are resummed.
In this contribution we outline a general approach \cite{Albino:2005gd} which 
extends the FO evolution of FFs to smaller $x$ values by resumming SGLs.
We present explicit results for the case where the 
largest SGLs, being the double logarithms (DLs), are resummed
using the Double Logarithmic Approximation
(DLA) \cite{Bassetto:1982ma;Fadin:1983aw,Dokshitzer:1991wu} in the LO DGLAP evolution,
and use this scheme to describe, and to fit FFs to, experimental data.
We also introduce a novel and simple approach to incorporate hadron mass effects.

\section{SGL Resummation in DGLAP Evolution}
\label{SGLDGLAP}

The DGLAP equation reads
\beq
\frac{d}{d\ln Q^2} D(x,Q^2)=\int_x^1 \frac{dy}{y}P(z,a_s(Q^2)) D\left(\frac{x}{y},Q^2\right),
\label{DGLAPx}
\eeq
where, for brevity, we omit hadron and parton labels. $D$ is a vector
containing the gluon FF $D_g$ and the quark and antiquark FFs $D_q$ and $D_{\overline{q}}$
respectively, in linear combinations according to the choice of basis, and
$P$ is the matrix of the splitting functions.
We define $a_s=\alpha_s/(2\pi)$.
We choose the simplest basis, consisting of valence quark FFs $D_q^-=D_q -D_{\bar{q}}$, 
non-singlet quark FFs $D_{NS}$, and $D=(D_{\Sigma},D_g)$, where for $n_f$
quark flavours the singlet quark FF is given by $D_{\Sigma}=\frac{1}{n_f}\sum_{q=1}^{n_f} D_q^+$.

Analytic operations are often simpler in Mellin space, where a function $f(x)$ becomes
$f(\omega)=\int_0^1 dx x^{\omega} f(x)$,
since the convolution in $x$ space in Eq.\ (\ref{DGLAPx}) becomes the simple product
\beq
\frac{d}{d\ln Q^2}D(\omega,Q^2)=P(\omega,a_s(Q^2))D(\omega,Q^2).
\label{DGLAPn}
\eeq

Consider the formal expansion of $P(a_s)$ in $a_s$
keeping $x$ (or $\omega$ if we want to work in Mellin space) fixed, viz.\
\beq
P(a_s)=\sum_{n=1}^{\infty}a_s^n P^{(n-1)}
\label{expanofPinaszfix}
\eeq
where the $P^{(n-1)}$ are functions of $x$ (or $\omega$).
Equation (\ref{expanofPinaszfix}) truncated at some chosen 
(finite) $n$ is known as the FO approach, and, in $x$ space, is not valid
at small $x$ due to the presence of terms which
in the limit $x\rightarrow 0$ behave like $(a_s^{n}/x) \ln^{2n-m-1}x$ for 
$m=1,...,2n-1$. Such logarithms are called SGLs, and $m$ labels their class. 
As $x$ decreases, these
unresummed SGLs will spoil the convergence of the FO series for $P(x,a_s)$ once $\ln (1/x) = O(a_s^{-1/2})$.
Consequently the evolution of $D(x,Q^2)$ will not be valid here, since
the whole range $x\leq y\leq 1$ contributes in Eq.\ (\ref{DGLAPx}). 

SGLs are defined to be all those terms 
of the form $a_s^n/\omega^{2n-m}$ only, where 
$m=1,...,2n$ and labels the class of the SGL, in the expansion about $\omega=0$
of Eq.\ (\ref{expanofPinaszfix}) in Mellin space. For $m=1,...,2n-1$,
this definition agrees with the form of the SGLs in $x$ space given above, since
$\omega^{-p}=-((-1)^p/p!)\int_0^1 dx x^{\omega}(1/x)\ln^{p-1} x$
for ${\rm Re} (\omega) >0$ and $p\geq 1$. 
Such terms spoil the convergence of the series in Eq.\ (\ref{expanofPinaszfix}) 
as $\omega \rightarrow 0$. To construct a scheme valid for large {\it and}
small $\omega$ (or $x$), we write $P$ in the form
$P=P^{\rm FO}+P^{\rm SGL}$, where $P^{\rm SGL}$ contains only and all the SGLs in $P$,
so that $P^{\rm FO}$ is completely free of SGLs. 
Since terms of the type
$p=0$ ($m=2n$), which are included in our definition of SGLs, are non-singular, they may therefore
be left unresummed. Thus we separate
$P^{\rm SGL}$ into $P^{\rm SGL}=P^{\rm SGL}_{p\geq 1}+P^{\rm SGL}_{p=0}$.
$P^{\rm SGL}_{p=0}$, which is independent of $\omega$, is expanded as a series in $a_s$.
On the other hand, by summing all
SGLs in each class $m$, $P^{\rm SGL}_{p\geq 1}(\omega,a_s)$ is resummed in the form
\beq
P^{\rm SGL}_{p\geq 1}(\omega,a_s)=\sum_{m=1}^{\infty}\left(\frac{a_s}{\omega}\right)^m
g_m \left(\frac{a_s}{\omega^2}\right),
\label{expanpsglinmel}
\eeq
and truncated for some finite $m$. 
The remaining FO contribution to $P$,
$P^{\rm FO}(\omega,a_s)$, is expanded in $a_s$ keeping $\omega$ fixed, 
$P^{\rm FO}(\omega,a_s)=\sum_{n=1}^{\infty}a_s^n P^{{\rm FO}(n-1)}(\omega)$,
and truncated for some finite $n$. 
Finally, the result for $P(\omega,a_s)$ is inverse Mellin transformed to obtain $P(x,a_s)$, 
and then Eq.\ (\ref{DGLAPx}) is solved exactly.
We shall call this the SGL+FO+FO$\delta$ scheme, where ``+FO$\delta$'' means that the $p=0$
terms, which are each proportional to $\delta (1-x)$ in $x$ space, are left as a FO
series in $a_s$. In summary,
this scheme is the result of resumming all SGLs for which $m=1,...,2n-1$ 
in the form of
Eq.\ (\ref{expanpsglinmel}), and treating all remaining terms as in the FO approach. 

In the region for which $a_s \ll 1$ and $x$ is above values for which
$\ln (1/x) = O(a_s^{-1/2})$, the SGL+FO(+FO$\delta$) scheme should give a good description
of the evolution for the following reasons. Firstly, Eq.\ (\ref{expanpsglinmel}) implies
$P^{\rm SGL}_{p\geq 1}(x,a_s)=\frac{1}{x\ln x}\sum_{m=1}^{\infty}\left(a_s \ln x\right)^m 
f_m \left( a_s \ln^2 x\right)$. Since $a_s \ln x$ is always small,
this series is a valid approximation when $x$ is small.
On the other hand, as $x\rightarrow 1$ the SGLs for the types $p\geq 1$ all vanish, and therefore
so does each term in the series. The full contribution from the type $p=0$ terms is just
$P^{\rm SGL}_{p=0}(x,a_s)=\delta(1-x)\sum_{n=1}^{\infty}C_n a_s^n$,
and the expansion of $P^{\rm FO}(x,a_s)$ in $a_s$,
$P^{\rm FO}(x,a_s)=\sum_{n=1}^{\infty}a_s^n P^{{\rm FO}(n-1)}(x)$,
is finite for all $x$.

\section{DLA Improved DGLAP Evolution}
\label{DLADGLAP}

From the DLA \cite{Dokshitzer:1991wu}, if the evolution is rewritten in the form
\beq
\begin{split}
\frac{d}{d \ln Q^2}D(x,Q^2)&=\int_x^1 \frac{dy}{y} \frac{2C_A}{y}A  y^{2\frac{d}{d\ln Q^2}}
\left[a_s(Q^2) D\left(\frac{x}{y},Q^2\right)\right]\\
&+\int_x^1 \frac{dy}{y}\overline{P}(y,a_s(Q^2)) D\left(\frac{x}{y},Q^2\right),
\end{split}
\label{DGLAPandDLA1}
\eeq
then $\overline{P}(x,a_s)$ is free of DLs.
Explicitly, $A=0$
for the DL evolving parts of the components $D=D_q^-$ and $D=D_{NS}$, while 
\begin{eqnarray}
A=\left( \begin{array}{cc}
0 & \frac{2 C_F}{C_A} \\
0 & 1
\end{array} \right)
\end{eqnarray}
for the component $D=(D_{\Sigma},D_g)$.
Note that $A$ is a projection operator, i.e.\ it obeys 
$A^2=A$. $\overline{P}$ can be determined from $P$ explicitly
by expanding the operator $y^{2\frac{d}{d\ln Q^2}}$ in Eq.\ (\ref{DGLAPandDLA1}) in 
$\frac{d}{d\ln Q^2}$ and using Eq.\ (\ref{DGLAPx}).

We will examine what constraint Eq.\ (\ref{DGLAPandDLA1})
provides for $P$. In Mellin space, Eq.\ (\ref{DGLAPandDLA1}) becomes
\beq
\begin{split}
\left(\omega+2\frac{d}{d \ln Q^2} \right)& \frac{d}{d \ln Q^2}D(\omega,Q^2)
=2C_A a_s(Q^2) A D(\omega,Q^2)\\
&\hspace{-0.6cm} +\left(\omega+2\frac{d}{d \ln Q^2}\right)\overline{P}(\omega,a_s(Q^2)) D(\omega,Q^2).
\end{split}
\label{DRAP}
\eeq
After substituting Eq.\ (\ref{DGLAPn}) into Eq.\ (\ref{DRAP}) and dividing out 
the overall factor of  $D(\omega,Q^2)$, we obtain
the following constraint on $P$:
\beq
\left(\omega+2\frac{d}{d\ln Q^2}\right)\left(P-\overline{P}\right)
+2\left(P-\overline{P}\right)P-2C_A a_s A=0.
\label{eqfordelP}
\eeq
We now use the fact that $\overline{P}$ is free of DLs, 
to obtain an explicit constraint for $P^{\rm DL}$.
We first make the replacement $P=\widetilde{P}+P^{\rm DL}$ in Eq.\ (\ref{eqfordelP}).
Expanding Eq.\ (\ref{eqfordelP}) as a series in $a_s/\omega$ keeping $a_s/\omega^2$ fixed and 
extracting the first, $O((a_s/\omega)^2)$, term gives
\beq
2(P^{\rm DL})^2+\omega P^{\rm DL}-2C_A a_s A=0.
\label{DLAeqsimplest}
\eeq
Equation (\ref{DLAeqsimplest}) gives two solutions for each component of $P$. Since
$P$ is never larger than a 2$\times$2 matrix in the basis consisting of
singlet, gluon, non-singlet and valence quark FFs, there are four solutions.
We choose the solution
\beq
P^{\rm DL}(\omega,a_s)=\frac{A}{4}\left(-\omega+\sqrt{\omega^2+16C_A a_s}\right),
\label{DLresummedinP}
\eeq
since, in the component $D=(D_{\Sigma},D_g)$, 
the expansion of the result in Eq.\ (\ref{DLresummedinP}) in 
$a_s$ to $O(a_s^2)$ keeping $\omega$ fixed,
\begin{eqnarray}
P^{\rm DL}(\omega,a_s)=
\left( \begin{array}{cc}
0 & a_s \frac{4 C_F}{\omega}-a_s^2 \frac{16 C_F C_A}{\omega^3} \\
0 & a_s \frac{2 C_A}{\omega}-a_s^2 \frac{8 C_A^2}{\omega^3} 
\end{array} \right)+O(a_s^3),
\label{NLODLinmelspace}
\end{eqnarray}
agrees with the DLs in the literature, while in the
components $D=D^-_q$ and $D=D_{NS}$, $P^{\rm DL}=0$.
The other possibilities 
do not give these results and/or cannot be expanded in $a_s$, i.e. they are
non-perturbative.
Equation (\ref{DLresummedinP}) agrees with previous results \cite{Dokshitzer:1991wu,Mueller:1982cq}.

At small $\omega$, Eq.\ (\ref{DLresummedinP}) implies that 
\beq
D_{q,\overline{q}} =\frac{C_F}{C_A}D_g,
\label{DLArelforDquarkandDg}
\eeq
obtained after integrating Eq.\ (\ref{DGLAPn}) and neglecting the constant,
which is valid at large $Q$. We will use Eq.\ (\ref{DLArelforDquarkandDg}) 
to partially constrain our choice of parameterization
at low $x$ in the next section.

The $O(a_s)$ single logarithm (SL) 
contribution to $P$ is a type $p=0$ term (see Sec.\ \ref{SGLDGLAP}), and is given by
\begin{eqnarray}
P^{{\rm SL}(0)}(\omega)=
\left( \begin{array}{cc}
0 & -3C_F \\
\frac{2}{3}T_R n_f \ & -\frac{11}{6}C_A-\frac {2}{3}T_R n_f
\end{array} \right),
\label{singlogsatLO}
\end{eqnarray} 
Approximating $\overline{P}$ by $a_s P^{{\rm SL}(0)}$ 
Eq.\ (\ref{DRAP}) can be regarded as a generalized version of the 
Modified Leading Logarithm Approximation (MLLA) 
\cite{Dokshitzer:1991wu,Mueller:1982cq,Dokshitzer:1984dx} equation to include quarks,
in the sense that the $g$ component of this latter equation for $D=(D_{\Sigma},D_g)$
when Eq.\ (\ref{DLArelforDquarkandDg}) is invoked is precisely
the MLLA equation. We therefore conclude Eq.\ (\ref{DRAP}) is more complete than the
MLLA equation.

$P^{\rm DL}(x,a_s)$ is obtained by
inverse Mellin transform of Eq.\ (\ref{DLresummedinP}), which yields
$P^{\rm DL}(x,a_s)=\frac{A\sqrt{C_A a_s}}{x\ln \frac{1}{x}}
J_1\left(4\sqrt{C_A a_s}\ln \frac{1}{x}\right)$,
where $J_1(y)$ is the Bessel function of the first kind, given by 
$J_1(y)=\frac{1}{\pi}\int_0^{\pi}d\theta \cos (y\sin \theta -\theta)$.

In the next section we shall consider the DL+LO+LO$\delta$ scheme, in which 
we take $P^{\rm SGL}\approx P^{\rm DL}$, 
and take $P^{\rm FO}$ (and $P^{\rm SGL}_{p=0}$) to $O(a_s)$ only. 

\begin{figure}[h!]
\centerline{\epsfxsize=7cm\epsfbox{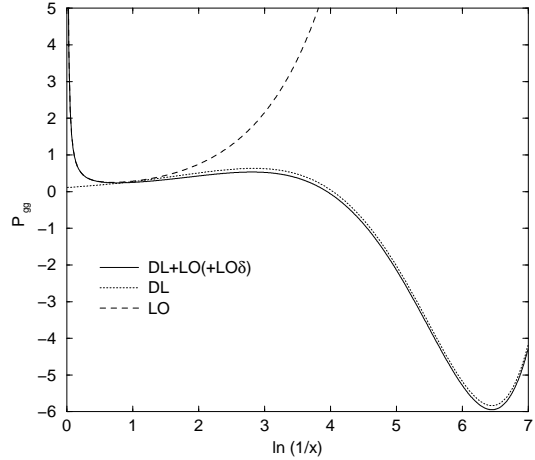}}
\caption{\label{plot1} (i) $P_{gg}(x,a_s)$ calculated in the DL+LO(+LO$\delta$) scheme,
(ii) $P_{gg}(x,a_s)$ calculated to $O(a_s)$ in the FO approach (labelled ``LO''), and 
(iii) $P_{gg}^{\rm DL}(x,a_s)$ (labelled ``DL''). $a_s=0.118/(2\pi)$.}
\end{figure}

In Fig.\ \ref{plot1}, we see that $P_{gg}(x,a_s)$  
in the DL+LO scheme, which is equal to the DL+LO+LO$\delta$ scheme when $x\neq 1$, 
interpolates well between its $O(a_s)$ approximation
in the FO approach at large $x$ and 
$P_{gg}^{\rm DL}(x,a_s)$ at small $x$ (the small difference here
comes from $P^{{\rm FO}(0)}(x)$ at small $x$). DL resummation clearly makes a 
large difference to $P$ at small $x$.

\section{Comparisons with data}
\label{CtD}

In this section, we study the outcome of fitting FFs evolved in the LO DGLAP
approach and in our DL+LO+LO$\delta$ scheme to experimental data on
the normalized differential cross section for light charged hadron production
in the process $e^+ e^- \rightarrow (\gamma,Z) \rightarrow h+X$, where $h$
is the observed hadron and $X$ is anything else. These data exist
for different center-of-mass (COM) energies $\sqrt{s}$ and values of
$x_p=2p/\sqrt{s}$, where $p$ is the
momentum of the observed hadron, which constrain the FFs in the region of $x$
for which $x_p\leq x \leq 1$. We fit to data for which
$\xi < \ln (\sqrt{s}/1 {\rm GeV})$, where $\xi=\ln (1/x_p)$.
At LO in the coefficient functions, these data are described in terms of the
evolved FFs by
\beq
\frac{1}{\sigma(s)}\frac{d\sigma}{dx_p}(x_p,s)=\frac{1}{n_f \langle Q(s) \rangle }
\sum_q Q_q(s) D_q^+(x_p,Q^2),
\label{approxxs}
\eeq
where $Q_q$ is the electroweak charge of a quark with flavour $q$ and $\langle Q \rangle$ is the average
charge over all flavours. We will take $n_f=5$ in all our calculations. 
Since we sum over hadron charges, we set $D_{\overline{q}}=D_q$.
These data depend on the FFs in the combinations
$f_{uc}(x,Q_0^2)=\frac{1}{2}\left(u(x,Q_0^2)+c(x,Q_0^2)\right)$,
$f_{dsb}(x,Q_0^2)=\frac{1}{3}\left(d(x,Q_0^2)+s(x,Q_0^2)+b(x,Q_0^2)\right)$
and the gluon $g(x,Q_0^2)$. For each of these three FFs, we choose the parameterization
\beq
f(x,Q_0^2)=N\exp[-c\ln^2 x]x^{\alpha} (1-x)^{\beta},
\label{genparam}
\eeq
since at intermediate and large $x$ the FF is constrained to behave like
$f(x,Q_0^2)\approx Nx^{\alpha} (1-x)^{\beta}$,
which is the standard parameterization used in global fits at large $x$, while at small $x$
(where $(1-x)^{\beta}\approx 1$) the FF is constrained to behave like
$\lim_{x\rightarrow 0}f(x,Q_0^2)=N\exp\left[-c\ln^2\frac{1}{x}-\alpha \ln \frac{1}{x}\right]$,
which for $c>0$ is a Gaussian in $\ln (1/x)$, as predicted 
by the DLA for sufficiently large $Q_0$.
We use Eq.\ (\ref{DLArelforDquarkandDg}) to remove four
free parameters by imposing the constraints
$c_{uc}=c_{dsb}=c_g$ and
$\alpha_{uc}=\alpha_{dsb}=\alpha_g$.
The relation
\beq
\begin{split}
N_{uc} \approx N_{dsb} \approx \frac{C_F}{C_A} N_g
\end{split}
\label{approxrelbetweenNs}
\eeq
implied by Eq.\ (\ref{DLArelforDquarkandDg}) is not imposed, since we want
to describe large $x$ data as well. We also fit $\Lambda_{\rm QCD}$.
We choose $Q^2=s$, although it is only important that the latter two quantities
are kept proportional, since the constant of proportionality has no effect on the final
FF parameters and the description of the data (and therefore the quality of the fit). 
This implies that there will be an overall
theoretical error on our fitted values for $\Lambda_{\rm QCD}$ of a factor of $O(1)$.
Since all data will be at $\sqrt{s}\geq 14$ GeV, we choose $Q_0=14$ GeV. 
The evolution is performed by numerically integrating
Eq.\ (\ref{DGLAPx}). 

\subsection{Fixed Order Evolution}
\begin{figure}[h!]
\centerline{\epsfxsize=7cm\epsfbox{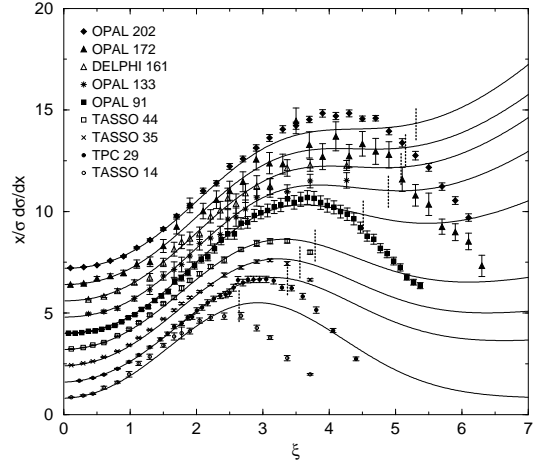}}
\caption{\label{fig1} Fit to data as described in Table \ref{tab1}. 
Some of the data sets used for the fit are shown,
together with their theoretical predictions from the results of the fit. Data to the right
of the vertical dotted lines have not been used in the fit. Each curve is shifted up by 0.8 for clarity.}
\end{figure}

We first perform a fit using standard LO DGLAP evolution.
We obtain $\chi^2_{\rm DF}=3.0$, and the results are shown in Fig.\ \ref{fig1} and
Table \ref{tab1}. The result for $\Lambda_{\rm QCD}$ is quite consistent with
that of other analyses, at least within the theoretical error.
It is clear that FO DGLAP evolution fails in the description of the peak region.
The unphysical negative value of $\beta$ for the gluon is 
because the gluon FF is weakly constrained,
since it couples to the data only through the evolution (see Eq.\ (\ref{approxxs})).
\begin{table}[h!]
\tbl{\label{tab1} Parameter values for the FFs at $Q_0=14$ GeV parameterized as in Eq.\
(\ref{genparam}) from a
fit to all data listed in the text using DGLAP evolution in the FO approach to LO.
$\Lambda_{\rm QCD}=388$ MeV.}
{\footnotesize
\begin{tabular}{c|llll}
\backslashbox{FF}{Parameter} & $N$  & $\beta$ & $\alpha$ & $c$  \\
\hline
                           $g$ & 0.22 & $-$0.43 & $-$2.38  & 0.25 \\
\hline
                         $u+c$ & 0.49 & 2.30    & [$-$2.38]       & [0.25]   \\
\hline
                       $d+s+b$ & 0.37 & 1.49    & [$-$2.38]       & [0.25]   
\end{tabular}}
\end{table}

\subsection{Incorporation of Soft Gluon Resummation}
\begin{figure}[h!]
\centerline{\epsfxsize=7cm\epsfbox{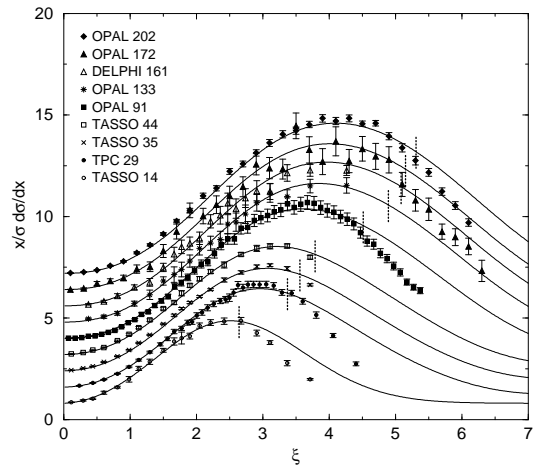}}
\caption{\label{fig2} Fit to data as described in Table \ref{tab2}.}
\end{figure}

We now redo the previous fit,
but now evolving in the DL+LO+LO$\delta$ scheme. 
The results are shown in Table \ref{tab2} and Fig.\ \ref{fig2}. 
\begin{table}[h!]
\tbl{\label{tab2} Parameter values for the FFs at $Q_0=14$ GeV parameterized as in Eq.\
(\ref{genparam}) from a fit to all data listed in the text 
using DGLAP evolution in the DL+LO+LO$\delta$ scheme.
$\Lambda_{\rm QCD}=801$ MeV.}
{\footnotesize
\begin{tabular}{c|llll}
\backslashbox{FF}{Parameter} & $N$  & $\beta$ & $\alpha$ & $c$  \\
\hline
                           $g$ & 1.60 & 5.01    & $-$2.63  & 0.35 \\
\hline
                         $u+c$ & 0.39 & 1.46    & [$-$2.63]       & [0.35]   \\
\hline
                       $d+s+b$ & 0.34 & 1.49    & [$-$2.63]       & [0.35]   
\end{tabular}}
\end{table}
We obtain $\chi^2_{\rm DF}=2.1$, a significant improvement to the fit
above with FO DGLAP evolution. In particular,
the data around the peak is now much better described.

We note that had we made the usual DLA (MLLA) choice
$Q=\sqrt{s}/2$ instead of our choice $Q=\sqrt{s}$ which is usually employed in analyses
using the DGLAP equation, we would have obtained half the result $\Lambda_{\rm QCD}\approx 800$ MeV.
$N_g$ is too large by a factor of about 2 relative to its prediction
in Eq.\ (\ref{approxrelbetweenNs}). However, as noted before, the initial gluon FF
is weakly constrained in our fits.

\subsection{Incorporation of Hadron Mass Effects}
To attempt to improve the description beyond the peak, we now study
hadron mass effects, which are important at small $x_p$.
It is helpful to work with light cone coordinates, 
in which any 4-vector $V$ is written in the form 
$V=(V^+,V^-,{\mathbf V_T})$ with $V^{\pm}=\frac{1}{\sqrt{2}}(V^0 \pm V^3)$ and
${\mathbf V_T}=(V^1,V^2)$. In the COM frame,
the momentum of the electroweak boson takes the form
$q=\left(\frac{\sqrt{s}}{\sqrt{2}},\frac{\sqrt{s}}{\sqrt{2}},
{\mathbf 0}\right)$. The momentum of the hadron in the COM frame is chosen as
$p_h=\left(\frac{\eta \sqrt{s}}{\sqrt{2}},\frac{m_h^2}{\sqrt{2}\eta \sqrt{s}},
{\mathbf 0}\right)$, where $\eta=p_h^+/q^+$ is the light cone scaling variable.
Therefore the relation between the two scaling variables
in the presence of hadron mass is
$x_p=\eta \left(1-\frac{m_h^2}{s\eta^2}\right)$.
As a generalization of the massless case, we assume the 
cross section we have been calculating is $(d\sigma/d\eta)(\eta,s)$, i.e.\ 
$\frac{d\sigma}{d\eta}(\eta,s)=\int_{\eta}^1 \frac{dy}{y} \frac{d\sigma}{dy}(y,s,Q^2)
D\left(\frac{\eta}{y},Q^2\right)$,
which is related to the measured observable $(d\sigma/dx_p)(x_p,s)$
via
\beq
\frac{d\sigma}{dx_p}(x_p,s)=\frac{1}{1+\frac{m_h^2}{s\eta^2(x_p)}}
\frac{d\sigma}{d\eta}(\eta(x_p),s).
\eeq

Although the data is for light charged hadrons, the vast majority of particles
are pions, so we will assume all particle masses are equal.
We now perform the DL+LO+LO$\delta$ fit again but with $m_h$ included in the list of free parameters.
We obtain the results in Table \ref{tab4}.
The parameters are not substantially different to those in Table \ref{tab2}.
The result for $m_h$ is reasonable for light charged hadrons.
We find $\chi^2_{\rm DF}=2.03$, i.e.\ no significant improvement to the
quality of the fit, and the comparison with data is similar to that in Fig.\ref{fig2}.
However, treatment of mass effects renders the value of $\Lambda_{\rm QCD}$
more reasonable.
\begin{table}[h!]
\tbl{\label{tab4} As in Table \ref{tab2}, but incorporating mass effects in the fit.
$\Lambda_{\rm QCD}=399$ MeV and $m_h=252$ MeV.}
{\footnotesize
\begin{tabular}{c|llll}
\backslashbox{FF}{Parameter} & $N$  & $\beta$ & $\alpha$ & $c$  \\
\hline
                           $g$ & 1.59 & 7.80 & $-$2.65  & 0.33 \\
\hline
                         $u+c$ & 0.62 & 1.43    & [$-$2.65]       & [0.33]   \\
\hline
                       $d+s+b$ & 0.74 & 1.60    & [$-$2.65]       & [0.33]   
\end{tabular}}
\end{table}

\section{Conclusions}
\label{conclusions}

Using the approach in this contribution gives a much better fit to all the data
than the FO approach and the MLLA \cite{Albino:2004xa} do, even
if the fit is still not in the acceptable range. 
Further improvement in the large $\xi$ region can be expected from the inclusion of higher order
SGLs. Our scheme allows a determination of quark and gluon FFs over a wider range of data
than previously achieved, and should be used to 
extend the NLO global fits of FFs to lower $x_p$ values.

\section*{Acknowledgments}
This work was supported in part by the Deutsche Forschungsgemeinschaft     
through Grant No.\ KN~365/5-1 and by the Bundesministerium f\"ur Bildung und  
Forschung through Grant No.\ 05~HT4GUA/4.



\end{document}